\documentclass[aps,epsf,preprint]{revtex4}
\usepackage{amssymb}

\def\1{{\bf 1}}
\def\[{\left[}
\def\]{\right]}
\def\be{\begin{eqnarray}}
\def\ee{\end{eqnarray}}
\def\bm{\begin{pmatrix}}
\def\em{\end{pmatrix}}
\def\nn{\nonumber}
\def\({\left(}
\def\){\right)}

\def\eq#1{(\ref{#1})}

\def\a{\alpha}

\def\s{\sigma}

\def\l{\lambda}

\def\x{\times}

\def\labels#1{\label{#1}}
\def\edc{\end{document}}

\def\bn{\begin{enumerate}}
\def\i{\item}
\def\en{\end{enumerate}}
\def\b{\beta}
\def\g{\gamma}

\def\ba{\begin{array}}
\def\ea{\end{array}}
\def\bc{\begin{center}}
\def\ec{\end{center}}

\def\edoc{\end{document}}

\def\^{$\wedge$}
\def\.{\!\cdot\!}
\def\nuh{{\nu\over 2}}
\begin{document}

\title{Permutation Symmetry of the Scattering Equations}
\author{C.S. Lam}
\address{Department of Physics, McGill University\\
 Montreal, Q.C., Canada H3A 2T8\\
and\\
Department of Physics and Astronomy, University of British Columbia,  Vancouver, BC, Canada V6T 1Z1 \\
Email: Lam@physics.mcgill.ca}

\begin{abstract}
Closed formulas for  tree amplitudes of $n$-particle scatterings of gluon, graviton, and massless scalar particles have been proposed by
Cachazo, He, and Yuan. It depends on $(n-3)$ quantities $\s_\a$ which satisfy a set of coupled {\it scattering equations}, with
momentum dot products as input coefficients. 
These equations are known to have $(n-3)!$ solutions, hence each $\s_\a$ is believed
to satisfy a single polynomial equation of degree $(n-3)!$. In this article, we derive the transformation properties of 
$\s_\a$ under momentum permutation, and verify them
with known solutions at low $n$, and with exact solutions at any $n$ for special momentum configurations. For momentum
configurations not invariant under a certain momentum permutation, new solutions can be obtained for the permuted 
configuration from these symmetry relations. 
These symmetry relations
for $\s_\a$  lead to  symmetry relations for the
$(n-3)!+1$ coefficients of the single-variable polynomials, whose correctness are checked with the known cases at low $n$. The extent to which
the coefficient symmetry relations can determine the coefficients is discussed.
\end{abstract}
\narrowtext
\maketitle
\section{Introduction}
The number of Feynman tree
diagrams for $n$-gluon scattering grows rapidly with $n$. There are 4 diagrams for $n=4$,\
25 diagrams for $n=5$, and 220 diagrams for $n=6$. By the time one gets to $n=12$, the number exceeds five billion.
It is therefore highly remarkable that if all the gluon helicities are the same, or only one of them
is different, the resulting tree amplitude sums up to be zero whatever $n$ is. If all but two are the same,
then the result of the sum consists of only one term, given by the Parke-Taylor formula \cite{PT}.  Recently, Cachazo, He, and Yuan (CHY) \cite{CHY}
were able to generalize
this formula to any helicity configuration, any spacetime dimension $D$, 
not only for gluon scattering, but also to graviton and massless scalar
scattering amplitudes. Later on similar expressions for the scattering of massive
scalar particles were also obtained \cite{DG1} .

The formula consists of a sum of $(n-3)!$ terms, each of which is associated with
a solution of the {\it scattering equations} 
\be \sum_{b\not=a}^n {k_a\.k_b\over\s_a-\s_b}=0\quad (a=1,2,\cdots,n)\labels{se}\ee
in the unknown variables $\s_a$, where $k_a$ are the incoming momenta. Though the number $(n-3)!$ is still very large  for large $n$, 
nevertheless it is very much smaller
than the number of Feynman diagrams. For example, these numbers are 1, 2, 6 respectively for $n=4,5,6$, to be compared with the numbers
4, 25, 220 of Feynman diagrams. For $n=12$, the number $9!=362,880$ is much smaller than 5 billion.

The scattering equations possess a M\"obius invariance, so three
of the $n$ variables $\s_\a$ can be fixed. By multiplying
through with the product of  denominators, the scattering equations can be turned into $n-3$ coupled
homogeneous polynomial equations of degree $n-3$ each. These equations are known
to have $(n-3)!$ solutions \cite{CHY}, hence the equation to determine a single variable $\s_\a$ is expected
to be a polynomial equation of degree $(n-3)!$.  It is generally not easy
to obtain even the single-variable polynomials, not to speak of their solutions. Though there is no problem to get
$n=4$ and $n=5$,  special technique is required for $n\ge 6$ \cite{SW, DG2}. However, there are special momentum configurations 
for which solutions can be
obtained for any $n$ \cite{DG2, CK}.

Analytic solution for $\s_\a$ is impossible to obtain except for low $n$, and for special momentum configurations. However, since
they are central to the CHY scattering formulas, it is desirable to find out as much about them as possible. In this article
we will discuss how $\s_\a$ transforms under momentum permutations.
 A set of {\it symmetry relations} is derived in the next section, and verified
 against the known solutions of $n=4$ and $n=5$. These relations will also be checked against the exact solutions in 
certain momentum configurations. 
 For momentum permutations which alter the special momentum configurations, new exact solutions can be obtained  for the altered
 configurations from these symmetry
relations.

Let $A_p\ (0\le p\le (n-3)!)$ be the coefficients of  the polynomial equation
 for a single $\s_\a$. Symmetry relations for $\s_\a$ lead to symmetry
relations for $A_p$ which will be worked out in Sec.~III. These  relations are checked using the known coefficients
for $n=4$, $n=5$, and $n=6$. In Sec.~IV, we discuss the amount of constraints put on $A_p$ just by their symmetry relations. The
cases for $n=4$ and $n=5$ are worked out in detail to illustrate the general discussion. Up to an overall normalization, the symmetry
relations determine all the four parameters controlling the $n=4$ equation, and  36 of the 45 parameters
involved in the $n=5$ single-particle polynomial equation.

\section{Permutation Symmetry}
Let us use M\"obius invariance to fix $\s_1=0,\ \s_2=1$, and $\s_n=\infty$. Then the scattering equations \eq{se}
for $\s_\a\ (3\le\a\le n-1)$ are
\be
0&=&k_1\.k_2+{k_1\.k_3\over\s_3}+{k_1\.k_4\over\s_4}+\cdots+{k_1\.k_{n-1}\over\s_{n-1}}\labels{se1}\\
0&=&k_2\.k_1+{k_2\.k_3\over 1-\s_3}+{k_2\.k_4\over 1-\s_4}+\cdots+{k_2\.k_{n-1}\over 1-\s_{n-1}}\labels{se2}\\
0&=&{k_3\.k_1\over\s_3}+{k_3\.k_2\over\s_3-1}+{k_3\.k_4\over\s_3-\s_4}+\cdots+{k_3\.k_{n-1}\over\s_3-\s_{n-1}}\labels{se3}\\
0&=&{k_4\.k_1\over\s_4}+{k_4\.k_2\over\s_4-1}+{k_4\.k_3\over\s_4-\s_3}+\cdots+{k_4\.k_{n-1}\over\s_4-\s_{n-1}}\labels{se4}\\
&&\hskip3cm \cdots\cdots\nn\\  
0&=&{k_{n-1}\.k_1\over\s_{n-1}}+{k_{n-1}\.k_2\over\s_{n-1}-1}+{k_{n-1}\.k_3\over\s_{n-1}-\s_3}+\cdots+{k_{n-1}\.k_{n-2}\over\s_{n-1}-\s_{n-2}}\labels{sen}
\ee

The solutions $\s_\a$ depend on $k_1,k_2,\cdots,k_n$, but we will skip the arguments and write
$\s_\a(k_1,k_2,\cdots,k_n)$ simply as $\s_\a$. If $s$ is a permutation of $n$ objects, then 
$\s_\a(k_{s(1)},k_{s(2)},\cdots,k_{s(n)})$ will be written as $\s_\a(s)$. In particular, if $s=(jk)$ is a  transposition, then we will also write $\s_\a(s)$ as $\s_\a(jk)$, rather than the more cumbersome notation $\s_\a((jk))$. 

\subsection{Symmetry relations}
The purpose of this subsection
is to obtain the following relations between $\s_\a(jk)$ and $\s_\b$\
($3\le\a,\b\le n-1$):
\be
\s_\a(12)&=&1-\s_\a\labels{12}\\
\s_\a(2\a)&=&{1\over\s_\a}\labels{2a}\\
\s_\a(2\b)&=& {\s_\a\over\s_\b}, \quad  (\b\not=\a),\labels{a2b}\ee
as well as
\be
\s_\a(1\a)&=&{\s_\a\over\s_\a-1}\labels{1a}\\
\s_\a(1\b)&=&{\s_\b-\s_\a\over\s_\b-1}, \quad  (\b\not=\a)\labels{a1b}\\
\s_\a(\a\b)&=&\s_\b, \quad  (\b\not=\a)\labels{aab}\\
\s_\a(\b\g)&=&\s_\a,  \quad  (\a,\b,\g\  {\rm different}),\labels{abc}\ee
and
\be
\s_\a(1n)&=&{1\over\s_\a}\labels{1n}\\
\s_\a(2n)&=&{\s_\a\over\s_\a-1}\labels{2n}\\
\s_\a(\a n)&=&1-\s_\a\labels{an}\\
\s_\a(\b n)&=&{\s_\a(1-\s_\b)\over\s_\a-\s_\b},\quad  (\b\not=\a).\labels{anb}\ee

To prove these results, let us start from \eq{se2}. Interchange $k_2$ with $k_1$ and compare the resulting
equation with \eq{se1},
we get \eq{12}. Next, exchange $k_3$ with $k_1$ in \eq{se1} and compare the result with \eq{se3}, we get
\eq{1a} and \eq{a1b} for $\a=3$. The proof for $\a>3$ is identical. Similarly, exchange $k_3$ with $k_2$ in
\eq{se2} and compare the result with \eq{se3}, we get \eq{2a} and \eq{a2b} for $\a=3$. Similar proof works for
$\a>3$ as well. Now interchange $k_4$ with $k_3$ in \eq{se3}, compare the result with \eq{se4}, and generalize
the result to other $\a, \b, \g$, we get \eq{aab} and \eq{abc}. 

The proof for relations involving $k_n$  is a little
more complicated, as $k_n$ does not appear  in the scattering equations above. We must use momentum
conservation to introduce $k_n$, and then go through procedures similar to those adopted above.
 For example, if we replace $k_1$ by $-\sum_{i=2}^n k_i$ in \eq{se2}, then it
becomes
\be
0=k_2.k_n+\sum_{\b=3}^{n-1}k_2\.k_\b\(1-{1\over 1-\s_\b}\)=k_2\.k_n+\sum_{\b=3}^{n-1}k_2\.k_\b{\s_\b\over\s_\b-1}.\ee
Now interchange $k_n$ with $k_1$ and compare the result with \eq{se2}, we get \eq{1n}. Similarly, replace $k_2$
in \eq{se1} by momentum conservation to get
\be
0=k_1\.k_n+\sum_{\a=3}^{n-1}\(1-{1\over\s_\a}\).\ee
Now interchange $k_n$ with $k_2$ and compare the result with \eq{se1}, we get \eq{2n}. Finally, fix a $\a\ge 3$,
and replace $k_\a$ in \eq{se1} using momentum conservation, then we get
\be
0={k_1\.k_n\over\s_\a}+\sum_{j=2,j\not=\a}^{n-1} k_1\.k_j\({1\over\s_\a}-{1\over\s_j}\),\ee
which can be written as
\be 0=k_1\.k_n+\sum_{j=2,j\not=\a}^{n-1} k_1\.k_j\(1-{\s_\a\over\s_j}\).\ee
Exchange $k_n$ with $k_\beta$ and compare with \eq{se1}. If we set $j=2$, we get \eq{an}. If we set
$j=\beta\not=\alpha$,  then we get \eq{anb} with $\a$ and $\b$ reversed. This completes the proof of
the symmetry relations \eq{12} to \eq{abc}.
\vskip.8cm
Here are some supplemental remarks about the symmetry equations.
\bn 
\i The symmetry relations are obtained by comparing equations that $\s_\a$ satisfy, not the solutions themselves.
As a result, the $\s_\a$ appearing on the right of these symmetry relations may be a different solution
than the $\s_\a$ appearing on the left.
\i
Since the permutation relation $(ij)=(ik)(kj)(ik)$ is true for any $i,j,k$, the symmetry relations 
obtained above are not all independent. For
momenta not involving $k_n$, we can derive \eq{1a} to \eq{abc} from
\eq{12} to \eq{a2b}. For relations involving $k_n$,
another relation \eq{1n} is needed to obtain everything else. Details are given in Appendix A.

\i It follows from \eq{12} and \eq{an}, \eq{2n} and \eq{1a}, \eq{1n} and \eq{2a} that
\be \s_\a=\s_\a((12)(\a n))=\s_\a((1n)(2\a))=\s_\a((2n)(1\a))\labels{klein}\ee
for every $3\le\a\le n-1 $. Since $(12)(\a n),\ (1n)(2\a),\ (2n)(1\a)$, together with the identity permutation, form
the Klein group $Z_2\x Z_2$, we shall  refer to this identity as the {\it Klein-group identity}.
\i If we know a single $\s_\a$, then  we can compute every other $\s_\b$ from \eq{aab}, so the hard work is to find the
solutions for a single $\a$.

\i Using momentum conservation, $k_n$ can be eliminated so $\s_\a$ may be considered as a function of 
$k_1, k_2, \cdots, k_{n-1}$. In that form, equations \eq{1n} to \eq{anb} will not be useful, so that the
only independent relations needed to be considered are \eq{12}, \eq{2a}, and \eq{a2b}. 

\i Equation \eq{a2b} differs from the other two fundamental symmetry relations in that  two different $\s$'s
appear on the right hand side. However, using \eq{aab}, we can get rid of $\s_\b$. Using also \eq{2a}, equation
\eq{a2b} can be written in the form
\be
\s_\a(2\a)\s_\a(2\b)\s_\a(\a\b)=1.\labels{cubic}\ee

\i With that, it is amusing to note that the three relations \eq{12}, \eq{2a}, \eq{cubic} can be written as linear, quadratic,
and cubic relations, respectively:
\be
\s_\a+\s_\a(12)&=&1,\nn\\
\s_a\s_a(2\a)&=&1,\nn\\
 \s_\a(2\a)\s_\a(2\b)\s_\a(\a\b)&=&1.\labels{123}\ee

\en

\subsection{Direct verification}
In this subsection the symmetry relations will be verified directly for
$n=4$ and $n=5$, where analytic solutions of the scattering equations are available. For four-dimensional spacetime,
$n=6$ solutions are also available \cite{SW}, but they are lengthy and their verification is not be carried out here. 
\subsubsection{\boldmath{$n=4$}}
There is only one nontrivial $\s_\a$, which is $\s_3$. It depends on the Mandelstam variables \footnote{In order
not to carry a factor of 2 around, in this paper we define all the (generalized) Mandelstam variables
with an additional factor of 1/2, to be $(\sum k_i)^2/2$.} $s=k_1\.k_2=k_3\.k_4,\ t=k_1\.k_3=k_2\.k_4$, and
$u=k_1\.k_4=k_2\.k_3$. These three variables are subject to the constraint $s+t+u=0$ because all  the particles
are massless.  The scattering equation obtained from \eq{se1} is 
\be s+{t\over \s_3}=0,\labels{sen3}\ee
whose solution is $\s_3=-t/s$. 

Under momentum permutation (12), $s\leftrightarrow s$ and $t\leftrightarrow u$. Thus $\s_3(12)=-u/s=(s+t)/s=1-\s_3$, verifying \eq{12}. Under momentum permutation (23), $u\leftrightarrow u$ and $s\leftrightarrow t$, so $\s_3(23)=-s/t=1/\s_3$, verifying \eq{2a}. Equation \eq{a2b} is irrelevant for $n=4$, thus all the independent
symmetry relations have been explicitly verified.

\subsubsection{\boldmath{$n=5$}}
$\s_3$ and $\s_4$ depend on six scalar products, $s=k_1\.k_2,\ t_1=k_1\.k_3,\ t_2=k_1\.k_4,\ u_1=k_2\.k_3,\
u_2=k_2\.k_4$, and $v=k_3\.k_4$. These six variables sum up to zero because $k_5^2=(k_1+k_2+k_3+k_4)^2=0$,
hence only five of them are independent. In what follows we shall take them to be $s, t_1, t_2, u_1, u_2$. Scalar products
involving $k_5$ can be obtained by momentum conservation.

After multiplying by the product of denominators, eqs.~\eq{se1} and \eq{se2} become
\be
0&=&s\s_3\s_4+t_1\s_4+t_2\s_3,\labels{s5a}\\
0&=&s(1-\s_3)(1-\s_4)+u_1(1-\s_4)+u_2(1-\s_3).\labels{s5b}\ee
A linear equation is obtained by subtracting the two,
\be
(s+t_1+u_1)\s_4+(s+t_2+u_2)\s_3=u_1+u_2,\labels{s5c}\ee
which can be used to eliminate either $\s_3$ or $\s_4$. Substituting the result back into \eq{s5a}, we get
a quadratic equation determining $\s_3$ or $\s_4$:
\be
0&=&s(s+t_2+u_2)\s_3^2+\(s(-s+t_1-t_2-u_1-u_2)+t_1u_2-t_2u_1\)\s_3-t_1(s+u_1+u_2)\nn\\
&:=&a\s_3^2+b\s_3+c,\labels{s5d}\\
0&=&s(s+t_1+u_1)\s_4^2+\(s(-s-t_1+t_2-u_1-u_2)-t_1u_2+t_2u_1\)\s_4-t_2(s+u_1+u_2)\nn\\
&:=&a'\s_4^2+b'\s_4+c',\labels{s5e}\ee
where
\be
a&=&s(s+t_2+u_2),\nn\\
b&=&s(-s+t_1-t_2-u_1-u_2)+t_1u_2-t_2u_1,\nn\\
c&=&-t_1(s+u_1+u_2),\nn\\
a'&=&s(s+t_1+u_1),\nn\\
b'&=&s(-s-t_1+t_2-u_1-u_2)-t_1u_2+t_2u_1,\nn\\
c'&=&-t_2(s+u_1+u_2).\labels{abcabc}\ee
The solutions of these quadratic equations are
\be
\s_3&=&(-b\pm\sqrt{b^2-4ac})/2a,\labels{s5f}\\
\s_4&=&(-b'\pm\sqrt{{b'}^2-4a'c'})/2a'.\labels{s5g}\ee
We will use $\s_{3+}$ and $\s_{3-}$ to denote  solution \eq{s5f} with the upper and the lower sign, and 
$\s_{4+}$ and $ \s_{4-}$ to denote solution \eq{s5g} with the upper and the lower sign.

With these explicit solutions, we are now ready to verify the  independent symmetry relations \eq{12} to 
\eq{a2b}. To do so, we need to know how the variables change under the momentum
permutation (12), (23), and (24). The result is listed in Table 1. Please remember that
$v=k_3\.k_4=-(s+t_1+t_2+u_1+u_2)$ is not independent, but its change is also listed in Table 1.
Also, the relations obtained from the permutation (34) is not independent either, but it will be convenient to list it as well.

$$\ba{|c|c|c|c|c|c|c|}\hline
&s&t_1&t_2&u_1&u_2&v\\ \hline
(12)&s&u_1&u_2&t_1&t_2&v\\
(23)&t_1&s&t_2&u_1&v&u_2\\
(24)&t_2&t_1&s&v&u_2&u_1\\ 
(34)&s&t_2&t_1&u_2&u_1&v\\
\hline
\ea$$
\vskip.5cm
\bc Table 1. Transformation of dot products under momentum permutations\ec
\vskip.7cm
Let us now consider each of the symmetry relations separately.
\vskip.5cm
\noindent\underline{\bf{Eq.~\eq{12}}}.\quad
Because of the opposite sign in front of the square roots for solutions $+$ and $-$ in \eq{s5f} and \eq{s5g}, clearly \eq{12} must be 
interpreted to mean
\be
\s_{\a \pm}(12)=1-\s_{\a \mp},\qquad (\a=3, 4).\ee
In order for that to be true, we must have
\be
a(12)=a,\quad b(12)=-(2a+b), \quad b(12)^2-4a(12)c(12)=b^2-4ac, \ee
and similarly with $a, b, c$ replaced by $a', b', c'$. These equalities are equivalent to
\be
a(12)=a,\quad b(12)=-(2a+b), \quad c(12)=a+b+c,\labels{abc12}\ee
and similarly with $a, b, c$ replaced by $a', b', c'$.
Using the explicit expressions in \eq{abcabc},
and the (12) row of Table 1, it is easily seen that these identities are true.

\vskip.5cm
\noindent\underline{\bf{Eq.~\eq{2a}}}.\quad Again, because of the opposite signs in front of the square roots,
we must interpret \eq{2a} to mean
\be
\s_{3\pm}(23)={1\over\s_{3\mp}},\quad \s_{4\pm}(23)={1\over\s_{4\mp}}.\ee
In order for those to be true, we must have
\be
a(23)&=&c,\quad b(23)=b, \quad c(23)=a\labels{abc23}\\
 a'(24)&=&c',\quad b'(24)=b, \quad c'(24)=a'.\ee
These relations can be verified explicitly from \eq{abcabc} and Table 1.

\vskip.5cm
\noindent\underline{\bf{Eq.~\eq{a2b}}}.\quad The verification of this is more complicated, because it involves
the ratio of two different $\s$'s. \eq{a2b} seems quite impossible unless the square roots in \eq{s5f} and \eq{s5g} are identical.
The expressions for $b^2-4ac$ and ${b'}^2-4a'c'$ are both rather lengthy, but straight-forward computation
shows that they are indeed equal. I shall use the letter $d$ to denote them.

Let us compute $\s_{3+}/\s_4$. A priori we do not know whether to use $\s_{4+}$ or $\s_{4-}$, but detailed 
calculation shows that it is $\s_{4-}$. In that case,
\be
{\s_{3+}\over\s_{4-}}={a'(-b+\sqrt{d})\over a(-b'-\sqrt{d})}={a'(-b+\sqrt{d})(-b'+\sqrt{d})\over a({b'}^2-d)}
={ d+bb'-(b'+ b)\sqrt{d}\over 4ac'}.
\ee
According to \eq{a2b}, this should be either $\s_{3+}(24)$ or $\s_{3-}(24)$. Again, detailed calculation shows
that it is the latter, namely,
$\s_{3-}(24)=\(-b(24)-\sqrt{d(24)}\)/2a(24)$. In order for that to be true,
it is necessary to have $d(24)=d$, which can be verified to be true. Moreover, we need to have
\be
-b(24)/a(24)=(d+bb')/2ac', \quad 1/a(24)=(b'+b)/2ac',\ee
or equivalently,
\be
a(24)=2ac'/(b+b'), \quad b(24)=-(d+bb')/(b+b').\labels{abcabcp}\ee
Explicit substitution shows that these are indeed true, hence \eq{a2b}, in the form $\s_{3-}(24)=\s_{3+}/\s_{4-}$,
is verified. Similarly, one can also verify $\s_{3+}(24)=\s_{3-}/\s_{4+}$  and $ \s_{4\mp}(23)=\s_{4\pm}/\s_{3\mp}$  to be true.

Equation \eq{a2b} is the most intriguing of the three  independent symmetry relations: it involves two different
$\s$'s on the right. As a result,  the symmetry constraints must involve parameters for both $\s_\a$ and
$\s_\b$. Furthermore, they must appear nonlinearly, as  in \eq{abcabcp}.

Clearly, for \eq{abcabcp} to be true, $a, b, c$ and $a', b', c'$ must be closely related. Indeed, equation \eq{aab}
tells us what their relations are. We will verify that directly below.
\vskip1cm
\noindent\underline{\bf{Eq.~\eq{aab}}}.\quad This equation says 
$\s_3(34)=\s_4,\  \s_4(34)=\s_3$. From the sign of the square roots, we now expect
$\s_{3\pm}(34)=\s_{4\pm}$, which demands
\be
a(34)=a',\quad  b(34)=b',\quad c(34)=c'.\ee
These relations can also be explicitly verified to be true.

This completes the verification of the symmetry relations for $n=5$.

\subsection{Special configurations}
For certain special momentum configurations, exact solutions for $\s_\a$ can be obtained for any $n$.
We discuss some of these found in the literature \cite{DG2, CK} in this subsection, to show that they either obey
the symmetry relations \eq{12}, \eq{2a}, and \eq{a2b}, or  these relations can be used to produce
exact solutions for permuted momentum configurations. In this connection please recall that the $\s$'s that appear
on the right of  \eq{12} to \eq{a2b} may be a different solution than what appears on the left.
\subsubsection{}
Consider the special configuration 
\be
 k_2\.k_\a&=&k_\a\. k_\b=\mu,\quad (3\le\a\not=\b\le n-1),\nn\\
k_1\.k_2&=&k_1\.k_\a=\nu,\labels{munu}\ee
where $\mu$ is arbitrary and $\nu$ is chosen so that $k_n^2=(\sum_{i=1}^{n-1}k_i)^2=0$, namely,
$\nu=-(n-3)\mu/2$. Dolan and Goddard \cite{DG2} showed that the solutions of the scattering equations are
\be
\s_2=1,\quad \s_\a=\omega_\a,\labels{munu2}\ee
where $\omega_\a$ are $(n-2)$th roots of unity, different for different $\a$, and none of them equal to unity. The $(n-3)!$ possibilities
of arranging such distinct $\omega_\a$'s constitute the $(n-3)!$ solutions of the scattering equations.

Symmetry relation \eq{2a} is obeyed because $\omega^{-1}_\a$ is another $(n-2)$th root of unity,  so it is just another
solution of $\s_\a$. Symmetry relation \eq{a2b} is also obeyed because $\s_\a/\s_\b$ is another $(n-2)$th root, not equal to unity if $\b\not=\a$.
Note that neither of the momentum permutations $(2\a)$ and $(2\b)$ changes the momentum configurations in \eq{munu}.

With momentum permutation (12), the configuration \eq{munu} changes into the configuration
\be
 k_1\.k_\a&=&k_\a\. k_\b=\mu,\qquad (3\le\a,\b\le n-1),\nn\\
k_1\.k_2&=&k_2\.k_\a=\nu,\labels{munu3}\ee
whose solutions, according to \eq{12}, are 
\be \s_2=1,\quad \s_\a=1-\omega_\a. \labels{munu4}\ee
This is a new solution for the new configuration \eq{munu3}.

\subsubsection{}
A more general configuration is considered by Kalousios \cite{CK}, in which $k_2\.k_\a$ is different from $k_\a\.k_\b$:
\be
k_1\.k_\a=(1+\nu)/2,\quad k_2\.k_\a=(1+\mu)/2, \quad k_\a\.k_\b=1,\qquad (3\le\a,\b\le n-1).\labels{CK}\ee
$k_1\.k_2$ is determined by the requirement of $k_n^2=0$ to be
\be
k_1\.k_2=-(n-3)(n+\mu+\nu-2)/2.\ee
According to Kalousios, the solution for $\s'_\a=2\s_\a-1$ is given by the $(n-3)$ roots of the Jacobi polynomial $P_{n-3}^{(\mu, \nu)}(\s'_\a)$.
In the $\s'$ space, these roots lie in the interval $[-1, 1]$, and they are known to reverse their sign when $\mu$ and $\nu$ are interchanged.
This property agrees with \eq{12}, which when expressed in the $\s'$ variable, says $\s'_\a(12)=-\s_\a'$. Furthermore, 
in this configuration, the momentum
permutation (12) is equivalent to the interchange of $\mu$ and $\nu$.

The momentum configuration \eq{CK} is {\it not} invariant under momentum permutation $(2\a)$. Thus this permutation produces
a new momentum configuration, and new solutions by using \eq{2a} and \eq{a2b}.

\section{Polynomial equations}
The scattering equations \eq{se} are equivalent to the polynomial equations  \cite{DG2} \footnote{$\s_i$ here is $1/z_i$ in
Ref.~\cite{DG2}. This inverts the role of $k_1$ and $k_n$ which is why the generalized Mandelstam variables
$s_{\cdots}$ here involve a $k_n$ rather than a $k_1$. We have also defined the Mandelstam variables
with an additional factor ${1\over 2}$ to avoid carrying the factor 2 everywhere in the dot products.}
\be  0=\sum_{a_1,a_2,\cdots,a_m\in A'} s_{a_1a_2\cdots a_m n}\s_{a_1}\s_{a_2}\cdots\s_{a_m}:=h_m,
\quad (1\le m\le n-3), \labels{pe}\ee
where $A'=\{2,3,4,\cdots,n-1\}$, $\s_2=1$, and $s_{a_1a_2\cdots a_m n}={1\over 2}(k_{a_1}+k_{a_2}+\cdots+k_{a_m}+k_n)^2$.
The sum is taken over all distinct subsets of $A'$ with $m$ elements. Using momentum conservation, we can get
rid of $k_n$ in favor of $k_1$ to write $s_{a_1a_2\cdots a_m n}$ as ${1\over 2}(k_1+k_{\bar a_1}+\cdots +k_{\bar a_{n-2-m}})^2$, where $\bar a_i$ are the complementary indices of $a_i$ in the set $A'$.

The advantage of these equations is that each $\s_i$ enters at most linearly. That makes it easier to eliminate all other variables to obtain a polynomial equation for a single variable. A single-variable polynomial can be easily
obtained directly from the scattering equations for $n=4$ and $n=5$, but using these polynomial equations, 
a sixth order single-variable polynomial is also derived for $n=6$ In Ref.~\cite{DG2}.

Since there are $(n-3)!$ solutions, it is natural to assume such single-variable polynomials  to be of degree $(n-3)!$.
Accordingly, let us write them in the form
\be 0=\sum_{p=0}^\nu A_p^{(\a)}\s^p_\a,\labels{pe1}\ee
where $A_p^{(\a)}$ is a function of $k_1,\cdots,k_{n-1}$, and $\nu=(n-3)!$. 

\subsection{Symmetry of \boldmath{$A_p^{(\a)}$}}
Under a permutation of the momenta $k_i$, $\s_\a$ transforms according to the symmetry relations \eq{12}, \eq{2a},  \eq{1a}, and \eq{abc}. 
They induce  corresponding relations between the polynomial coefficients $A_p^{(\a)}$ which we will work out in this subsection. The other symmetry relations involves another variable $\s_\b$ so they will not 
be immediately useful in determining the symmetry relations of $A_p^{(\a)}$.

Under $\s\to 1-\s$, \eq{pe1} changes into $0=\sum_{p=0}^\nu\tilde A_p\s^p$, with
\be  (-)^{p-\nu}\tilde A_p&=&\sum_{q=p}^{\nu}{q!\over p!(q-p)!}A_q\nn\\
&=&A_p+(p+1)A_{p+1}+{(p+1)(p+2)\over 2!}A_{p+2}+\cdots+{(p+1)(p+2)\cdots \nu\over(\nu-p)!}A_\nu,\labels{pe2}\ee
where an overall constant $(-)^\nu$ has been inserted to make $\tilde A_\nu=A_\nu$.

Under $\s\to 1/\s$, \eq{pe1} changes into  $0=\sum_{p=0}^\nu \bar A_p\s^p$, with
\be \hat A_p=A_{\nu-p}.\labels{pe3}\ee

Under $\s\to \s/(\s-1)$, \eq{pe1} changes into $0=\sum_{p=0}^\nu \overline A_p\s^p$, with
\be (-)^{\nu-p}\bar A_p&=&\sum_{q=0}^p{(\nu-p+q)!\over q!(\nu-p)!}A_{p-q}\nn\\
&=&A_p+(\nu+1-p)A_{p-1}+{(\nu+1-p)(\nu+2-p)\over 2!}A_{p-2}+\cdots\nn\\
&&+ {(\nu+1-p)(\nu+2-p)\cdots\nu\over p!} A_0.\ee

Therefore \eq{12}, \eq{2a}, and \eq{1a}  translate into the relations
\be A_p^{(\a)}(12)=\tilde A_p^{(\a)},\quad   A_p^{(\a)}(2\a)=\hat A_p^{(\a)},\quad  A_p^{(\a)}(1\a)=\bar A_p^{(\a)}.\labels{aprel}\ee
The third one can be derived from the first two so it will be ignored from now on.
In addition, \eq{aab} and \eq{abc} imply
\be A^{(\a)}_p(\b\g)&=&A_p^{(\a)},\labels{aprel2}\\
A^{(\a)}_p(\a\b)&=&A^{(\b)}_p.\labels{aprel3} \ee

These relations will be verified directly for $n=4,5,6$.

\subsection{Direct verification for  \boldmath{$n=4, 5, 6$}}
\subsubsection{\boldmath{$n=4$}}
The single-variable linear polynomial is given in \eq{sen3}, with
$A_1=s$ and $A_0=t$. Thus
\be
A_1(12)=s=A_1,\quad A_0(12)=u=-(s+t)=-(A_1+A_0),\ee 
which agrees with \eq{aprel}. 

\subsubsection{\boldmath{$n=5$}}
Equation \eq{aprel} requires
\be
A_2(12)=A_2,\quad A_1(12)=-(A_1+2A_2),\quad A_0(12)=A_0+A_1+A_2.\ee
In the notation of \eq{s5d}, $A_2=a, A_1=b, A_0=c$, so this relation is simply \eq{abc12}, which has been verified.
Similarly we can verify it for $\s_4$. It also requires
\be
A_2(23)=A_0,\quad A_1(23)=A_1, \quad A_0(23)=A_2,\ee
which is just \eq{abc23} and that has also been verified. Similarly, one can verify the
identities for $\s_4$ as well.

\subsubsection{\boldmath{$n=6$}}
Single-variable polynomial equations have been obtained for $n=6$ \cite{SW, DG2}.
Following the recipe given in eq.~(3.9) of Ref.~\cite{DG2},
with $x=\s_3$ and $y=\s_4$ in that equation, a sixth degree polynomial for $\s_5$ can be obtained. Its $A_0$
and $A_6$ coefficients are given by
\be
A_0&=&(k_4\.k_5+k_3\.k_5+k_4\.k_3+k_1\.k_5+k_1\.k_4+k_1\.k_3)^2(k_4\.k_5+k_1\.k_5+k_1\.k_4)\x\nn\\
&&(k_1\.k_5)^2(k_3\.k_5+k_1\.k_5+k_1\.k_3),\nn\\
A_6&=&(k_4\.k_3+k_4\.k_2+k_2\.k_3+k_1\.k_4+k_1\.k_3+k_1\.k_2)^2(k_4\.k_2+k_1\.k_4+k_1\.k_2)\x\nn\\
&&(k_1\.k_2)^2(k_2\.k_3+k_1\.k_3+k_1\.k_2),\labels{n6}\ee
and the other coefficients are much too long to write down. According to \eq{aprel}, we should have
$A_6(12)=A_6$, and $A_6(25)=A_0$, which can be seen from \eq{n6} to be true.

\section{Symmetry constraints}
We found the symmetry relations for the coefficients $A_p^{(\a)}$ of the single-variable polynomials
in the last section. In this section we investigate the inverse, and ask to what
extent the symmetry relations determine the coefficients $A_p^{(\a)}$, and hence the single-variable
equations. 

Note that the symmetry relations \eq{aprel} for $A_p^{(\a)}$, derived from \eq{12} and \eq{2a}, are weaker than
the symmetry relations for $\s_\a$, as equation \eq{a2b} was not used. Since \eq{a2b}
involves two $\s$'s, it is difficult to translate it into equations for $A_p^{(\a)}$. In the special case $n=5$, 
we do  know what
it is. It is given by \eq{abcabcp}, which is nonlinear and very complicated, but we do not know how to generalize
that to a larger $n$. As a result, the symmetry relations for $A_p^{(\a)}$ are certainly `incomplete', so we
do not expect it to be able to yield the complete expressions for $A_p^{(\a)}$, save for the case $n=4$
where \eq{a2b} is irrelevant. Still, we would like to see how far they can take us.

More precisely, pretend that we do not
 know about the scattering equations, either in the original form \eq{se},  or
in its polynomial form \eq{pe}, but we assume that we do know the following. Each $\s_\a$ satisfies a $\nu=(n-3)!$th 
degree polynomial equation, whose coefficients $A_p^{(\a)}$ are 
homogeneous polynomials of the scalar products
$k_i\.k_j\  (1\le i, j\le n-1) $, also of degree $\nu$. Moreover, $A_p^{(\a)}$ satisfy the symmetry relations \eq{aprel} and \eq{aprel2}. 
The task is to find out to what extent these relations determine $A_p^{(\a)}$.

For most of this section we will be discussing a single $\a$, so this script will be dropped. To be definite, we will take $\a=3$, though the treatment is identical for the other
$\a$'s.

The symmetry equations under permutation (12) are
\be
+A_\nu(12)&=&A_\nu,\nn\\
-A_{\nu-1}(12)&=&A_{v-1}+\nu A_\nu,\nn\\
+A_{\nu-2}(12)&=&A_{\nu-2}+(\nu-1)A_{\nu-1}+{1\over 2!}(\nu-1)\nu A_\nu,\nn\\
-A_{\nu-3}(12)&=&A_{\nu-3}+(\nu-2)A_{\nu-2}+{1\over 2!}(\nu-2)(\nu-1)A_{\nu-1}+{1\over 3!}(\nu-2)(\nu-1)\nu A_\nu,\nn\\
&&\hskip1cm\cdots\nn\\
(-)^\nu A_0(12)&=&A_0+A_1+\cdots +A_\nu.\labels{12exp}\ee
Let  $A_p=B_p+C_p$, where $B_p$ is even under permutation (12) and $C_p$ is odd. Then \eq{12exp}
is equivalent to the following set of more transparent equations,
\be
C_\nu&=&0,\nn\\
-2B_{\nu-1}&=&\nu B_\nu,\nn\\
2C_{\nu-2}&=&(\nu-1)C_{\nu-1},\nn\\
-2B_{\nu-3}&=&(\nu-2)B_{\nu-2}+{1\over 2!}(\nu-2)(\nu-1)B_{\nu-1}+{1\over 3!}(\nu-2)(\nu-1)\nu B_\nu,\labels{12exp2}\ee
etc. In other words, $C_\nu=0$,  but $B_\nu, C_{\nu-1}, B_{\nu-2}, C_{\nu-3}, \cdots$ are arbitrary. Moreover, the
equations give alternately constraints on $B$ and on the next $C$.

The remaining symmetry relations state that $A_p(23)=A_{\nu-p}$.  To see how they can be exploited,
we need to know the explicit forms for $B_p$ and $C_p$ as a function of $k_i\.k_j$.

In a  scattering process involving $n$ massless particles, the number of
 of scalar products 
that can be constructed from momenta $k_1,\cdots,k_{n-1}$ is $(n-1)(n-2)/2$. Since $0=k_n^2=(\sum_{i=1}^{n-1}k_i)^2$, these scalar products sum up to zero, so there are
only $\mu:=(n-1)(n-2)/2-1=n(n-3)/2$ different ones.  The functions
$A_p$ are $\nu$th degree homogeneous polynomials in these $\mu$ variables. As such, it contains
$\lambda:=(\nu+\mu-1)!/(\mu-1)!\nu!$ terms, and requires $\l$ coefficients to fix. Since there are
$(\nu+1)$ $A_p$'s, the total number of parameters needed to determine the coefficients $A_p$ of a single-variable polynomial
is $\kappa:=(\nu+1)\l$. For $n=4, 5, 6$, we have respectively $\nu=1, 2, 6$,\ $\mu=2, 5, 9$,\ $\l=2, 15, 3003$, and $\kappa=4, 45, 21021$. One of these parameters is an overall normalization constant which can never be determined.
The question we ask is how many of these $\kappa$ parameters can be determined by the symmetry relations \eq{aprel}, and how to calculate them.

We will describe the general procedure to be followed for any $n$, then proceed to carry out the 
explicit calculations for $n=4$ and $n=5$.
For $n=4$, up to normalization, all the $\kappa=4$ parameters can be determined. 
For $n=5$, out of the $\kappa=45$ parameters, 36 can be determined, leaving behind 9 arbitrary parameters.

To discuss the general procedure, 
it is more convenient to use variables that are either even or odd under the momentum exchange (12). 
The variables $k_1.k_i+k_2\.k_i$, for $3\le i\le n-1$, are even under the exchange, 
the variables $k_1.k_i-k_2\.k_i$ are odd, and
all the other scalar products are even. We shall use $x_i$ to denote these odd/even variables. Since $B_p$
is even under momentum permutation (12), it must contain an even number of the odd-variables in every term.
Similarly $C_p$ must contain an odd number of the odd-variables in every term.

For $n\ge 5$, $\nu$ is even, so $\nuh$ is an integer. That allows us to start from the
most convenient relation, $A_{\nuh}(23)=A_{\nuh}$, which is self conjugate.  
Let $x_i$ become $x_i'=\sum_{j=1}^\mu M_{ij}x_j$ under the momentum exchange (23). It is easy to work out 
what the matrix $M$ is for any $n$. The relation $A_{\nuh}(x_i')=A_{\nuh}(x_i)$
therefore provides $\l$ linear equations to determine the $\l$ coefficients of 
the $x$-monomials in $A_{\nuh}$. These linear equations
are homogeneous, so an overall normalization can never be determined. In addition, these equations may not be independent, so the solution may contain arbitrary parameters. In the case $n=5$, out of 15 parameters,
6 are determined, leaving behind 9 free ones.

Once $A_{\nuh}$, and hence $B_{\nuh}$ and $C_{\nuh}$, are known, \eq{12exp} can be used to relate it to $A_{\nuh+1}$ and $A_{\nuh-1}$,
then the relation $A_{\nuh+1}(23)=A_{\nuh-1}$ provides further linear equations to determine more unknown
parameters, and so on. Continuing this way, we can find out about all the parameters.

We proceed now to illustrate this general procedure with the specific examples $n=4$ and $n=5$.

\subsection{\boldmath$n=4$}
For $n=4$, the Mandelstam variables $s=k_1\.k_2=k_3\.k_4,\
 t=k_1\.k_3=k_2\.k_4$, $u=k_1\cdot k_4=k_2\.k_3$ add up to zero, so there is only one even variable $x_1=s$, and
one odd variable $x_2=t-u$. The functions  $A_1=p_1x_1+q_1x_2$ 
and $A_0=p_0x_1+q_0x_2$ are determined by four parameters, $p_1, q_1, p _0, q_0$. 
Symmetry relations \eq{12exp2} tell us  $q_1=0$ and $2p_0=-p_1$. 
Under (23) permutation, $x_1=s\to t=(x_2-x_1)/2=x_1'$ and $x_2=t-u\to s-u=(3x_1+x_2)/2=x_2'$. The relation
$A_1(23)=A_0$ gives rise to $p_1x_1'=p_1(x_2-x_1)/2=p_0x_1+q_0x_2$, thus $2q_0=p_1$. 
Consequently, $A_1=p_1x_1=p_1s$ and $A_0=p_1(-x_1+x_2)/2=p_1t$. The scattering equation is then 
completely determined by the symmetry relations to be  $A_1\s_3+A_0=0=s\s_3+t$, 
the same as equation \eq{sen3}.

\subsection{\boldmath{$n=5$}}
The single-variable polynomial is of degree 2. Each $A_p$ is quadratic in the 5 Mandelstam variables defined in Sec.~IIB, with $\l=15$ terms and hence 15 coefficients for each of $p=0,1,2$. The even/odd variables are
$ x_1=s,\ x_2=t_1+u_1,\ x_3=t_2+u_2,\ x_4=t_1-u_1,\ x_5=t_2-u_2$, with the first
three even and the latter two odd.  Together they form a column vector $x=(x_1,x_2,x_3,x_4,x_5)^T$.
Let $\xi_p$  be  $5\x 5$ symmetric
matrices whose elements are the unknown coefficients, so that $A_p=x^T\xi_px$. Parametrize $\xi_1$ as follows,
\be
\xi_1=\bm{p_{11}&p_{12}&p_{13}&q_{11}&q_{12}\cr p_{12}&p_{22}&p_{23}&q_{21}&q_{22}\cr
p_{13}&p_{23}&p_{33}&q_{31}&q_{32}\cr q_{11}&q_{21}&q_{31}&r_{11}&r_{12}\cr
q_{12}&q_{22}&q_{32}&r_{12}&r_{22}\cr}\em,\labels{xi1mat}\ee
then $p_{ij}$ and $r_{ij}$  contribute to $B_1$ and  $q_{ij}$ contribute to $C_1$.

From Table 1 we see that under a (23) momentum permutation,  $x\to x'=Mx$, with
{\footnotesize\be
M={1\over 2}\bm{0&1&0&1&0\cr 2&1&0&-1&0\cr -2&-2&-1&0&1\cr 2&-1&0&1&0\cr 2&2&3&0&1\cr}\em.
\labels{B23a}\ee}
The relation $A_1(23)=A_1$ is equivalent to $M^T\xi_1M=\xi_1$. This gives rise to a set of 15 linear equations,
whose solution leaves 9 parameters  ($r_{11}, r_{12}, r_{22}, q_{11}, q_{21}, q_{22}, q_{31}, q_{32}, p_{22}$)
free. The rest are given by
\be
p_{11}&=&2 q_{21}-4 q_{31}+4 r_{12}+p_{22}+r_{11},\nn\\
p_{12}&=&p_{22}-2 q_{31}+2 r_{12}+q_{11}-r_{11},\nn\\
p_{13}&=&-q_{31}+2 r_{12}+q_{22}-2 q_{32}+2 r_{22},\nn\\
p_{23}&=&q_{31}-r_{12}+q_{22}-2 q_{32}+2 r_{22},\nn\\
p_{33}&=&-2 q_{32}+3 r_{22},\nn\\
q_{12}&=&q_{31}+q_{22}.\labels{xi10}\ee
Using \eq{12exp}, we get
\be \xi_2=-\bm{p_{11}&p_{12}&p_{13}&0&0\cr p_{12}&p_{22}&p_{23}&0&0\cr
p_{13}&p_{23}&p_{33}&0&0\cr 0&0&0&r_{11}&r_{12}\cr
0&0&0&r_{12}&r_{22}\cr}\em,\ee
and from $A_0=A_2(23)$, we can compute
\be
\xi_0=M^T\xi_2M.\ee
Thus, the 45  coefficients a priori need to determine $A_2, A_1, A_0$ are now reduced to only 9 arbitrary coefficients.
The single-variable polynomial equation derived from the scattering equation is given in \eq{s5d}. It corresponds
to the special case
\be
(r_{11}, r_{12}, r_{22}, q_{11}, q_{21}, q_{22}, q_{31}, q_{32}, p_{22})=\(0, 0, 0, {1\over 2}, 0, -{1\over 4}, {1\over 4}, 0,  0\).\labels{explicit}\ee

\section{Summary}
The $n$-particle tree amplitudes for gauge, gravity, and scalar scatterings are given by the CHY formulas, 
in terms of  external momenta $k_i$ and the 
solutions $\s_i (1\le i\le n)$ of their scattering equation. 
The scattering equation is M\"obius invariant, so three of the $\s_i$'s can be fixed, to be $\s_1=0,\ \s_2=1$, 
and $\s_n=\infty$
here. Each of the remaining $\s_\a$ satisfies a polynomial equation of degree $(n-3)!$, hence analytic solutions can
be obtained only for $n=4$ and $n=5$. Nevertheless, certain general properties of $\s_\a$ can be worked out for all $n$. We
have discussed in this article the momentum dependence of $\s_\a$, specifically, how it changes under a momentum permutation.
Since $\s_1, \s_2, \s_n$ are special, the momentum dependence of $\s_\a$ on $k_1, k_2, k_n$ are also special. 
The result is given in eqs.~(7) to (17), and is explicitly verified for $n=4$ and $n=5$ where analytic solutions 
are available. We have also discussed to what extent these symmetry properties determine the solutions $\s_\a$,
and how they lead to analogous transformation properties of the coefficiens of the polynomial equations for $\s_\a$.

Knowing the transformation of $\s_\a$, one should be able to work out how
the various pieces of the CHY scattering formulas behave under momentum permutations. Such an investigation is underway.

\appendix

\section{Group property of the symmetry relations}
Here are the details how \eq{1a} to \eq{abc} can be derived from \eq{12} to \eq{a2b}:
\be
\s_\a(1\a)&=&\s_\a((12)(2\a)(12))=1-\s_\a((2\a)(12))=1-{1\over\s_\a(12)}={\s_\a\over\s_\a-1}\nn\\
\s_\a(1\b)&=&\s_\a((12)(2\b)(12))=1-\s_\a((2\b)(12))=1-{\s_\a(12)\over\s_\b(12)}={\s_\a-\s_\b\over 1-\s_\b}\nn\\
\s_\a(\a\b)&=&\s_\a((2\a)(2\b)(2\a))={1\over\s_\a((2\b)(2\a))}={\s_\b(2\a)\over\s_\a(2\a)}=\s_\b\nn\\
\s_\a(\b\g)&=&\s_\a((2\b)(2\g)(2\b))={\s_\a((2\g)(2\b))\over\s_\b((2\g)(2\b))}={\s_\a(2\b)\over\s_\b(2\b)}=\s_\a.\nn
\ee
We need in addition \eq{1n} to show \eq{2n} to \eq{anb}:
\be
\s_\a(2n)&=&\s_\a((12)(1n)(12))=1-\s_\a((1n)(12))=1-{1\over\s_\a(12)}={\s_\a\over\s_\a-1}\nn\\
\s_\a(\a n)&=&\s_\a((2\a)(2n)(2\a))={1\over\s_\a((2n)(2\a))}={\s_\a(2\a)-1\over\s_\a(2\a)}=1-\s_\a\nn\\
\s_\a(\b n)&=&\s_\a((2\b)(2n)(2\b))={\s_\a((2n)(2\b))\over\s_\b((2n)(2\b))}={\s_\a(2\b)/(\s_\a(2\b)-1)
\over\s_\b(2\b)/(\s_\b(2\b)-1)}\nn\\
&=&{\s_\a/(\s_\a-\s_\b)\over 1/(1-\s_\b)}={\s_\a(1-\s_\b)\over\s_\a-\s_\b}.\nn
\ee


\end{document}